\documentclass{PoS}

\newcommand\FigWidth{.6\textwidth}
\newcommand\Eq[1]{Eq.~\ref{eq:#1}}
\newcommand\Fig[1]{Fig.~\ref{fig:#1}}
\newcommand\Sec[1]{Sec.~\ref{sec:#1}}
\newcommand\bfn{\mathbf n}
\newcommand\bfe{\mathbf e}
\newcommand\calB{\mathcal B}
\newcommand\calC{\mathcal C}
\newcommand\calO{\mathcal O}

\title{Avoiding the boson sign problem at finite chemical potential}

\ShortTitle{Avoiding the boson sign problem at finite chemical potential}

\author{
\speaker{Michael G. Endres}\\
Institute for Nuclear Theory, University of Washington, Seattle, WA 98195-1550\\
E-mail: \email{endres@u.washington.edu}
}

\abstract{
The usual path integral formulation for scalar particles at finite density involves a sign problem, making numerical simulation impractical.
We present alternative methods free of this difficulty.
We apply these approaches to $|\phi|^4$ theory in $2+1$ dimensions, presenting preliminary numerical results for physical quantities in the vicinity of the phase transition in the $T-\mu$ plane.
}

\FullConference{XXIVth International Symposium on Lattice Field Theory\\
                 July 23-28, 2006\\
                 Tucson, Arizona, USA}

\begin{document}

\section{Introduction}
\label{sec:1}
A large class of fermionic and bosonic lattice theories exhibit severe sign problems at finite density, rendering numerical attempts to study such theories impractical.
Attention has focused largely on fermionic theories and lattice quantum chromodynamics in particular, spurring numerous studies at chemical potentials $\mu<T$.
In this regime, the sign problem becomes less severe and can be dealt with by a variety of techniques.
\footnote{See \cite{Philipsen:2005mj} and references therein.}
At present, a generic solution to the sign problem that is valid in all regimes remains to be found.
We take a less ambitious step toward this end, focusing on relativistic scalar field theories that exhibit sign problems.
We sketch two related methods for solving the sign problem in the case of $SO(N)$ models at finite temperature and density, and use these methods to study the phase diagram when $N=2$ and in $2+1$ dimensions.
A detailed description of the methods, algorithms and results can be found in a future publication.

We begin by considering a $U(1) \simeq SO(2)$ scalar field theory at finite chemical potential.
Generalization of our results to the global symmetry group $SO(N)$ will be briefly discussed in \Sec{2}.
We work on an $L^d \times \beta$ lattice (in lattice units, where $a=1$) and impose periodic boundary conditions in the space and imaginary-time directions.
The lattice action for this model, defined on Euclidean space-time, is given by
\begin{eqnarray}
\label{eq:action}
S[\phi] = \sum_\bfn \left[\sum_\nu \left( 2|\phi_\bfn|^2 - \phi^\ast_\bfn e^{-\mu\delta_{\nu,0}} \phi_{\bfn+\bfe_\nu}
          - \phi^\ast_{\bfn+\bfe_\nu} e^{\mu \delta_{\nu,0}} \phi_\bfn \right) + V(|\phi_\bfn|) \right]\ .
\end{eqnarray}
The chemical potential $\mu$ has been introduced as a constant and imaginary zero-component of a gauge field, following the prescription of \cite{Hasenfratz:1983ba}.
At finite chemical potential, the action specified by \Eq{action} is complex.
As a result, standard Monte Carlo techniques are inapplicable because the quantity $e^{-S[\phi]}$ no longer has an interpretation as a probability weight for generating statistical ensembles.
We demonstrate how to avoid this issue by finding alternative representations for both the partition function and expectation values of operators in question.

\section{Alternative representations for $Z(\mu)$}
\label{sec:2}
We present two representations for the partition function, derived from strong coupling expansion techniques, that avoid the sign problem.
The first representation is based on the character expansion--an approach used in the more familiar context of lattice gauge theory.
\footnote{
For a review of these methods, see \cite{Ukawa:1979yv,Creutz:1984mg}.
}
After expressing the complex scalar field in terms of radial and angular modes, one expands the weight $e^{-S[\phi]}$ in terms of the characters of $U(1)$.
Subsequently, it is possible to integrate out the angular degrees of freedom.
The resulting partition function takes the schematic form
\begin{eqnarray}
\label{eq:Z1}
Z_1(\mu) = \sum_{\{l_{\bfn,\nu}\}\in{\mathbb Z}} \int_0^\infty [\rho d\rho]\, e^{-\tilde S_1[\rho,l] + \beta \mu Q[l]}\ ,
\end{eqnarray}
where the sum over integer-valued link fields $l$ represents the sum over characters (Fourier modes) and $\rho$ is the radial degree of freedom.
The link field $l$ represents the conserved current associated with the $U(1)$ symmetry of this model.
As such, the sum over link fields is constrained to the subspace of physical configurations which satisfy the divergence-free constraint
\begin{eqnarray}
\label{eq:divfree}
\sum_\nu \left( l_{\bfn,\nu} + l_{\bfn,-\nu} \right) = 0\ ,
\end{eqnarray}
where $l_{\bfn,-\nu} = -l_{\bfn-\bfe_\nu,\nu}$.
In this representation, the chemical potential couples to the charge $Q$, which is a functional of the link field $l$ and given by
\begin{eqnarray}
\label{eq:charge}
\beta Q[l] = \sum_\bfn l_{\bfn,0}\ .
\end{eqnarray}
The action $\tilde S_1$ is a complicated functional of the conserved current and radial mode.
Nonetheless, one can verify that the action is real and therefore the partition function is given by a sum over real and positive weights, as desired for the purpose of Monte Carlo simulations.

A second viable representation for the partition function is obtained from a hopping parameter expansion.
Here, one Taylor expands the exponential of the nearest neighbor part of the action and then integrates term by term.
Each term in the expansion is characterized by two link field configurations $m$ and $n$.
These link field configurations simply count the number of instances the nearest neighbor interaction and its complex conjugate appear in the expansion.
Consequently, the link fields take on non-negative integer values.
The partition function takes the form
\begin{eqnarray}
\label{eq:Z2}
Z_2(\mu) = \sum_{\{m_{\bfn,\nu}\}\in{\mathbb N}} \sum_{\{n_{\bfn,\nu}\}\in{\mathbb N}} \, e^{-\tilde S_2[m,n] + \beta \mu Q[m-n]}\ ,
\end{eqnarray}
where the action $\tilde S_2$ is real and, once again, the chemical potential couples to the total charge $Q$.
In this representation, the charge is a functional of the difference $m-n$ and thus we identify this quantity as the conserved current associated with the $U(1)$ symmetry of our theory.
One can explicitly verify that the difference $m-n$ satisfies the divergence-free constraint given by \Eq{divfree}.

The methods described above easily generalize to the case of $SO(N)$ theories with scalar fields transforming in the fundamental representation.
Recall that the chemical potential is an anti-symmetric, real-valued matrix living in the adjoint representation of $SO(N)$.
With an appropriate change of basis, it is possible to block-diagonalize the chemical potential, where each block is a real $2 \times 2$ anti-symmetric matrix.
If $N$ is even, the theory may be regarded as a theory of $N/2$ complex scalar fields coupled to $N/2$ independent chemical potentials and the aforementioned methods apply individually to each of the charged scalar fields.
If $N$ is odd, an additional real and neutral scalar field will be present.

\section{Algorithms}
\label{sec:3}
\begin{figure}[t]
\centering
\includegraphics[width=\FigWidth]{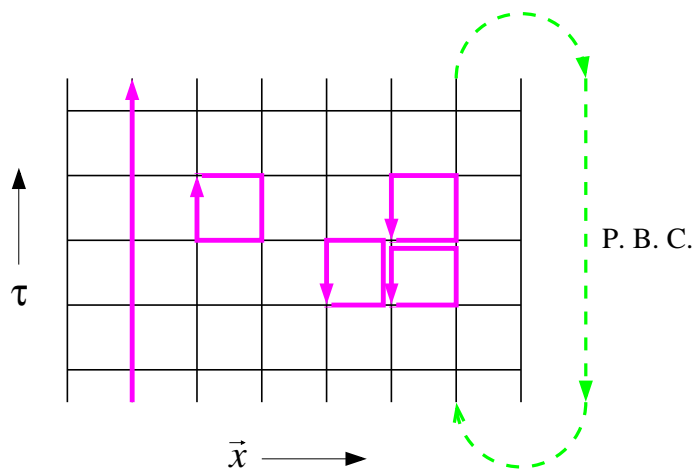}
\caption{Examples of global (left) and local (center) current loops. Arbitrary configurations (right) are obtained by multiple updates of local loops.} 
\label{fig:loops}
\end{figure}

We employ two algorithms for the purpose of simulating the partition functions $Z_1$ and $Z_2$.
The first algorithm is applied to $Z_1$ and is based on the Metropolis \cite{Metropolis:1953am} accept/reject method.
The main challenge of this approach is to perform updating of link variables in such a way as to maintain the constraint given by \Eq{divfree}.
The problem is easily circumvented if one updates simultaneously link variables that form closed loops (conserved current loops) as shown in \Fig{loops}; loops may either be local--forming the boundary of a plaquette, or global--forming a loop around the space-time torus.
The former is formally equivalent to solving the constraint by a change of variables to integer valued plaquettes \cite{Ukawa:1979yv}.
The chemical potential couples to fluctuations in the global current loops and consequently these fluctuations occur with a probability enhanced by a factor $e^{\beta \mu}$.
Though we have not addressed the question here, it is possible to calculate expectation values of all neutral observables with this algorithm.

We use the ``worm'' algorithm \cite{Prokofev:2001gh} to perform numerical simulations of the partition function $Z_2$;
physical link field configurations are generated by the motion of a dynamical source (or sink) throughout the space-time lattice.
The source traverses and consequently updates bond states with a probability governed by detailed balance.
In this algorithm, only a subset of the configurations that are generated are physical; these correspond to closed loop configurations which arise when the source and sink occupy the same site.
At finite chemical potential, the probability for forward motion of the source in the imaginary-time direction is enhanced by the factor $e^{\mu}$.
In a sequence of updates, it is possible and likely that the dynamical source will wind about the space-time torus resulting in an increase (or decrease) in the charge of the system at finite chemical potential.
The worm algorithm provides a simple method for obtaining expectation values of link field dependent operators as well as the unnormalized distribution of $2$-point correlation functions.
While limited in utility in some respects, the algorithm is efficient in that it takes advantage of unphysical ``fractional charge'' intermediate states to undergo charge fluctuations, thereby avoiding tunneling processes that are inherent to the loop algorithm previously described.

\section{An application}
\label{sec:4}
We apply the above formalism to a $2+1$ dimensional complex scalar field theory at finite temperature and density with the potential $V = m^2 |\phi|^2+ \frac{\lambda}{4} |\phi|^4$.
Numerical simulations of this model are performed using both the loop and worm algorithms with the aim of understanding the phase diagram in the $T-\mu$ plane and in the vicinity of the phase transition.
We present preliminary results of the effort.

\begin{figure}[t]
\centering
\includegraphics[width=\FigWidth]{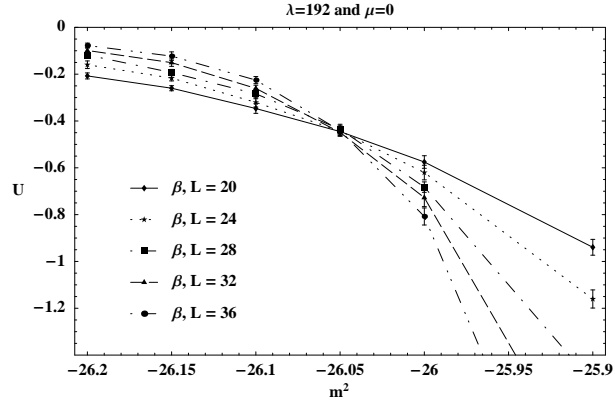}
\caption{$U$ as a function of $m^2$ for $\beta, L = 20-36$.} 
\label{fig:UvsMsq}
\end{figure}

At zero temperature and density, the theory under consideration is expected to exhibit a second order phase transition, located by first fixing the self-coupling at $\lambda=192$ and then tuning the mass parameter to the critical value $m_c^2$.
The critical value for the mass parameter is determined by studying the Binder cumulant
\begin{eqnarray}
\label{eq:U}
U(m^2) = 1- \frac{\langle Q^4 \rangle}{3\langle Q^2 \rangle^2}
\end{eqnarray}
as a function of $m^2$.
At large finite volumes, the charge distribution is Gaussian when $m \lesssim m_c$ and exponential when $m \gtrsim m_c$.
Consequently, in the thermodynamic limit $U(m^2)$ will approach zero in the broken phase and diverge in the unbroken phase.
Results obtained via the worm algorithm are displayed in \Fig{UvsMsq} for linear space-time volumes $\beta, L = 20 - 36$.
We estimate the critical mass from the intersection of level curves at large space-time volumes, finding $m_c^2 \approx -26.05$ for $\lambda=192$.

\begin{figure}[t]
\centering
\includegraphics[width=\FigWidth]{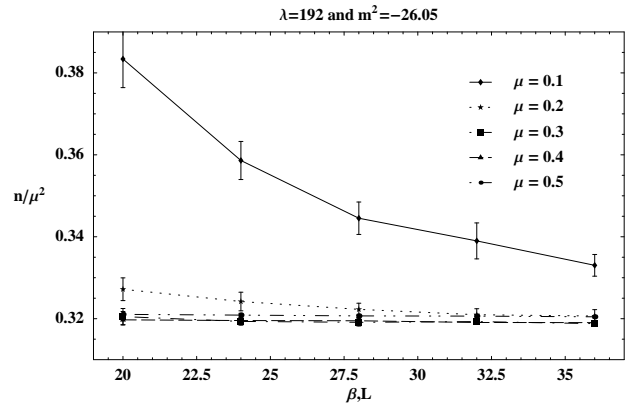}
\caption{$n/\mu^2$ as a function of $\beta, L$ for $\mu = 0.1-0.5$.} 
\label{fig:NvsL}
\end{figure}
Near the critical point $(m_c^2, \lambda)$ we investigate the dependence of the number density as a function of chemical potential.
According to \cite{Son:2002uz}, the number density should depend quadratically on the chemical potential and the curvature $\calB$ is expected to be a universal and nonperturbative constant.
We determine the curvature by plotting the ratio $n/\mu^2$, where $n$ is the number density, as a function of the linear space-time volume for values of the chemical potential ranging from $\mu=0.1-0.5$.
Results shown in \Fig{NvsL} were once more obtained using the worm algorithm.
For $\mu>0.1$, the level curves converge to $\calB \approx 0.32$ in apparent agreement with an indirect determination of $\calB =0.32(1)$ by \cite{Neuhaus:2002fp}.

\begin{figure}[t]
\centering
\includegraphics[width=\FigWidth]{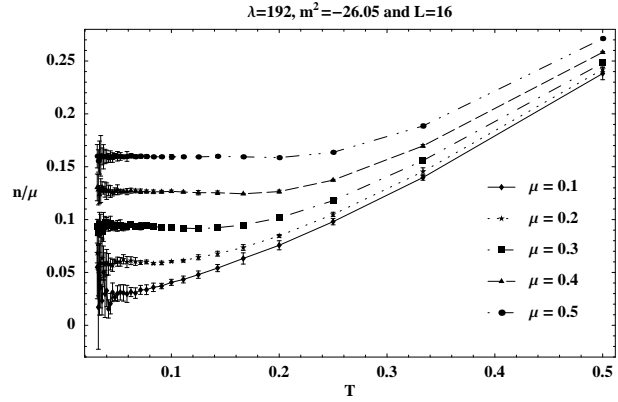}
\caption{$n/\mu$ as a function of temperature for $\mu=0.1-0.5$.} 
\label{fig:NvsT}
\end{figure}

Using the loop algorithm, we study the qualitative behavior of the number density as a function of temperature for chemical potentials ranging from $\mu=0.1-0.5$.
At temperatures $T \ll \mu$, the effective theory for the Goldstone mode obtained by the techniques of \cite{Nishida:2005uf} predicts weak dependence of the number density on temperature.
Specifically, one finds
\begin{eqnarray}
\label{eq:n}
n(\mu) = \calB \mu^2 \left[1 + \calO\left( T/\mu \right)^4 \right]\ ,
\end{eqnarray}
which appears in agreement with our numerical calculations displayed in \Fig{NvsT}.
\begin{figure}[t]
\centering
\includegraphics[width=\FigWidth]{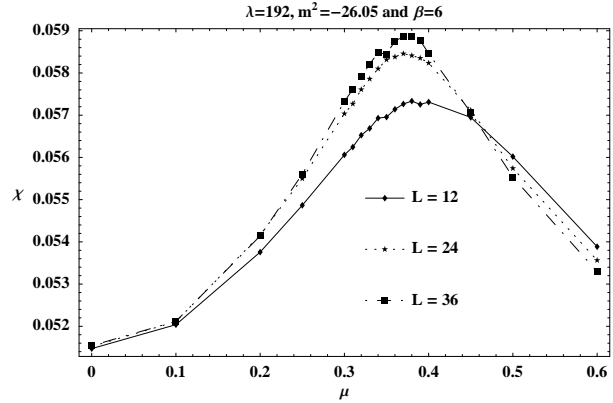}
\caption{$\chi$ as a function of $\mu$ for $L=12-36$.} 
\label{fig:ChiVsMu}
\end{figure}
Furthermore, we expect the shift in the critical temperature to depend linearly on the chemical potential \cite{Son:2006rt}, where the constant of proportionality $\calC$ is universal and nonperturbative. 
We can determine the value of $\calC$ by studying the chemical potential dependence of the susceptibility associated with $|\phi|^2$ for fixed $\beta$.
The susceptibility is defined by
\begin{eqnarray}
\label{eq:chi}
\chi = \beta L^d \left( \langle \overline{|\phi|^2}^2 \rangle - \langle \overline{|\phi|^2} \rangle \langle \overline{|\phi|^2} \rangle  \right)\ ,
\end{eqnarray}
where $\overline{|\phi|^2}$ is the space-time volume average of $|\phi|^2$.
An estimate of $\calC$ is attainable from the location of the maxima of the susceptibility $\mu_m(L)$ as one approaches the thermodynamic limit $L \to \infty$.
In this limit, the susceptibility is expected to diverge at the phase transition and therefore we expect $\calC^{-1} = \beta \mu_m(\infty)$.
Preliminary results for the susceptibility at linear spatial volumes $L=12-36$ and inverse temperature $\beta=6$ are shown in \Fig{ChiVsMu}.
A precise determination of $\calC$ is in progress.

\section{Conclusion}
\label{sec:5}
We have presented two alternative representations for the partition function for scalar field theories at finite density and temperature.
These representations are free of sign problems and are applicable to the numerical study of $SO(N)$ models with fundamental matter.
We have outlined several algorithms for simulating these partition functions and have applied the methods to $|\phi|^4$ theory in $2+1$ dimensions.
Preliminary numerical results for the universal constant $\calB$ appear in agreement with previous studies and our finite temperature results are qualitatively consistent with analytic calculation.

It is unlikely that the expansion techniques we have discussed are applicable to fermionic theories exhibiting sign problems.
Nonetheless, we believe that the pursuit of other representations for fermionic theories is worth consideration.

\section{Acknowledgments}
\label{sec:6}
M. G. E. would like to thank D. B. Kaplan, D. T. Son and M. Wingate for helpful suggestions throughout the course of this work.
This work was supported by U. S. Department of Energy grants DE-FG02-00ER41132.
\bibliography{bosons}
\bibliographystyle{h-physrev.bst}

\end{document}